\documentclass[
    ,final            
  ]
  {aipproc}
\usepackage{axodraw}
\layoutstyle{6x9}

\begin{document}

\title{4D Effective Theory and Geometrical Approach}

\classification{11.25.Mj, 04.50.+h.} \keywords {Extra dimensions,
Kaluza-Klein modes}

\author{A. Salvio}{
  address={SISSA and INFN, Via Beirut 2-4, 34014 Trieste,
Italy. } }

\begin{abstract}
 We consider the 4D effective theory for the light Kaluza-Klein (KK)
modes. The heavy KK mode contribution is generally needed to
reproduce the correct physical predictions: an equivalence, between
the effective theory and the D-dimensional (or {\it geometrical})
approach to spontaneous symmetry breaking (SSB), emerges only if the
heavy mode contribution is taken into account. This happens even if
the heavy mode masses are at the Planck scale. In particular, we
analyze a 6D Einstein-Maxwell model coupled to a charged scalar and
fermions. Moreover, we briefly review non-Abelian and supersymmetric
extensions of this theory.
\end{abstract}

\maketitle

\section{Introduction}

The low energy limit of a higher dimensional theory is usually
studied by taking into account only the light mode contribution. The
masses are derived from the bilinear part of the effective action
and the role of the heavy modes in the actual values of the masses
and the couplings of the effective theory for the light modes are
seldom taken into account. However, we know that the process of
"integrating out" the heavy modes \cite{Wilson} has the effect of
modifying the couplings of the light modes or introducing additional
terms that are suppressed by inverse powers of the heavy masses
\cite{Appelquist:1974tg}.

In a first part of this contribution we will summarize the study of
the heavy mode contribution to the low energy dynamics of higher
dimensional models, performed in Ref. \cite{Randjbar-Daemi:2006gf}.
There two methods have been used. The first one, which is called the
{\it 4D effective theory approach}, starts from a solution of a
higher dimensional theory and develops an action functional for the
light modes. This effective action generally has a local symmetry
that should be broken by Higgs mechanism. Our interest is in the
broken phase of the effective theory. The procedure is essentially
what is adopted in the effective description of higher dimensional
theories including superstring and M-theory compactifications. In
this construction the heavy KK modes are generally ignored simply by
reasoning that their masses are of the order of the compactification
mass and this can be as heavy as the Planck mass.

In the second approach, which we shall call the $\it{ geometrical\
approach}$, we shall find a solution of the higher dimensional
equations with the same symmetry group as the one of the broken
phase of the 4D effective theory. We shall then study the physics of
the light modes around this solution. The result for the low energy
physics will turn out to be different from the first approach.  The
difference is precisely due to the fact that in constructing the
effective theory along the lines of the first approach the
contribution of the heavy KK modes has been ignored.

This statement has been explicitly proved, at the classical level,
for a quite general higher dimensional scalar model with lagrangian
${\cal L}=-\frac{1}{2}\partial_M\Phi\partial^M\Phi+V(\Phi)$, where
$\Phi$ is a set of scalar fields, and then extended to a more
interesting (Abelian) gauge and gravitational theory
\cite{Randjbar-Daemi:2006gf}. Here we will briefly report the latter
case, which, in the low energy limit, reduces to a framework that is
similar to the electroweak part of the Standard Model. In this
framework, the heavy KK mode contribution can be geometrically
interpreted as the deformation of the internal space.

Moreover, in a final section, we shall review possible extensions of
these results to non-Abelian theories without fundamental scalars or
to supersymmetric versions of 6D gauge and gravitational theories.
In the former framework the Higgs field is identified with the
internal components of the non-Abelian gauge field
\cite{Dvali:2001qr} and the complete 6D gauge symmetry relaxes the
dependence of the Higgs mass on the ultraviolet cutoff. The latter
class of theories can be used as toy models for string theory
compactifications \cite{toykklt} and has shown some promise in
addressing the cosmological constant problem \cite{Burgess:2004ib}.

\section{6D Einstein-Maxwell-Scalar Model}

Here we analyze a 6D model, which includes the Einstein-Hilbert
gravity, a Maxwell field $A$ and a complex charged scalar $\phi$.
The bosonic action reads\footnote{Our conventions are
$\eta_{MN}=(-1,+1,+1,+1,\dots)$ and $R_{MN}=\partial_P \Gamma_{MN}^P
-\partial_M \Gamma_{PN}^P +\dots$.}
$$ S_B=\int\, d^6X
\sqrt{-G}\left[\frac{1}{\kappa^2}R-\frac{1}{4}F^2 -|\nabla \phi|^2
-V(\phi)  \right].$$
 where $F=dA$, $\nabla \phi=(d+ieA)\phi$. We choose $V(\phi)=m^2|\phi|^2+\xi|\phi|^4+\lambda$,
where $m^2$ is a real mass squared, $\xi$ is a real and positive
parameter and $\lambda$ represents a 6D cosmological constant. This
system is a simple generalization of the 6D Einstein-Maxwell model
of Ref. \cite{RSS}, where it was proved that the space-time
$(Minkowsky)_4\times S^2$ is a solution of the equations of motion
(EOMs), in the presence of a monopole background. This solution is
\begin{eqnarray}ds^2 &=&\eta_{\mu
\nu}dx^{\mu}dx^{\nu}+a^2\left(d\theta^2+\sin^2\theta
d\varphi^2\right),\nonumber\\ A&=&\frac{n}{2e}(cos\theta \mp 1)d\varphi,\label{solution1} \\
\phi&=&0, \nonumber\end{eqnarray}
where $a$ is the radius of $S^2$ and $n$ is the monopole number
($n=0,\pm 1,\dots$). Besides 4D Poincar\'e invariance, this
background preserves an $SU(2)\times U(1)$ symmetry, which turns out
to be the gauge symmetry of the low energy 4D effective theory
\cite{RSS}. The group factor $SU(2)$ has a geometrical origin as the
isometry group of the internal space, whereas $U(1)$ represents the
bulk gauge symmetry.

Moreover it is possible to introduce a couple of fermions
$\psi_{\pm}$, with 6D chirality $\pm 1$, and standard Yukawa
couplings: $\mathcal{L}_{Yuk}= g_Y \overline{\psi_+}\psi_-
\phi^{\dagger}+ g_Y\phi \overline{\psi_-}\psi_+$, where $g_{Y}$ is
assumed to be real for simplicity. Furthermore we assume that the
corresponding $U(1)$ fermion  charges are $e_+=e/2$ and $e_-=3e/2$,
because this corresponds to a simple fermion harmonic expansion over
$S^2$.

A complete study of the fluctuations around Solution
(\ref{solution1}) shows that the low energy 4D bosonic spectrum
presents the following states: the graviton, the $SU(2)\times U(1)$
gauge fields and a complex scalar field $\chi$, coming from $\phi$,
in the $(|n|+1)$-dimensional representation of $SU(2)$. For example,
for $n=2$ we obtain a triplet and, henceforth, we will assume $n=2$,
as the geometrical approach turns out to be simple in this case.
Moreover, in the fermion sector, we have a right-handed singlet and
a left-handed triplet.
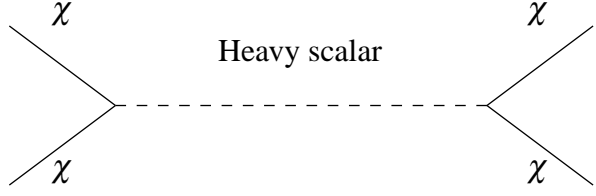
\begin{figure}
\begin{picture}(255,120)(0,20)

\DashLine(50,80)(190,80){4}\Text(120,100)[]{Heavy scalar}
\Line(190,80)(230,50)\Text(210,55)[]{$\chi$}
\Line(190,80)(230,110)\Text(210,115)[]{$\chi$}
\Line(10,50)(50,80)\Text(30,55)[]{$\chi$}
\Line(10,110)(50,80)\Text(30,115)[]{$\chi$}

\end{picture}
\caption{Heavy mode contribution to the quartic couplings. Reprinted
from \protect\cite{Randjbar-Daemi:2006gf}.}\label{diagram}
\end{figure}

\paragraph{4D Higgs Mechanism}

If the gauge symmetry is unbroken, the states that we have mentioned
above are exactly massless, apart from $\chi$. Indeed we can
introduce a small\footnote{It is small in the sense that
$\mu<<1/a$.} mass $\mu$ for $\chi$ by choosing a suitable value of
$m$. In order to create a small mass for fermions and gauge fields,
usually one computes an action functional for the light modes,
including bilinear terms and interactions, and then studies the
Higgs mechanism in the corresponding 4D theory. In our case, this
can be achieved by generalizing the zero-mode ansatz method of Ref.
\cite{RSS}, to include the light scalar $\chi$. In particular, the
lagrangian for $\chi$ turns out to be of the form ${\cal
L}_{\chi}=-(D_{\mu}\chi)^{\dagger}D^{\mu}\chi-\mathcal{U}(\chi),$
where $D_{\mu}$ is the $SU(2)\times U(1)$ covariant derivative and
${\cal U}$ is the scalar potential for $\chi$. In our model the
$U(1)$ and $SU(2)$ gauge constants, which appear in $D_{\mu}$, are
respectively given by
\begin{eqnarray} g_1=\frac{e}{\sqrt{4\pi}\,a},\quad
g_2=\sqrt{\frac{3}{16\pi}}\,\frac{\kappa}{a^2},\label{g12}\end{eqnarray}
whereas the potential is $\mathcal{U}(\chi)=\mu^2 \chi^{\dag} \chi
+\lambda_1(\chi^{\dag}\chi)^2 +\lambda_2|\chi^T \chi |^2+\dots,$
where the dots represent higher order operators and the $\lambda_i$
are the quartic coupling constants allowed by the $SU(2)\times U(1)$
gauge symmetry. We have $\lambda_1= \lambda_H+c_1\lambda_G,$ and
$\lambda_2= -(\lambda_H+c_2\lambda_{G})/3,$ where
$$\lambda_H=\frac{9}{20 \pi a^2}\xi,\quad \quad \lambda_G=
\frac{9\kappa^2}{80\pi a^4}.$$ We observe that $\lambda_{H}$ and
$\lambda_G$ represent respectively the light mode and the heavy mode
contribution to $\lambda_i$. The constants $c_i$ parametrize the
latter contribution, and can be explicitly computed by evaluating
diagrams of the form given in Fig. \ref{diagram}.

Now we focus on the SSB of $SU(2)\times U(1)$ down to $U(1)$, and we
assume $c_i=0$. In this phase ($\mu^2<0$), $\chi$ acquires a
non-vanishing vacuum expectation value (VEV), which, for $c_i=0$, is
given by
\begin{equation} |<\chi>|^2=\frac{3}{4}\,\,\frac{-\mu^2}{\lambda_H}. \label{VEV1}\end{equation}
The corresponding vector, fermion and scalar spectrum, at the
leading non trivial order in $\sqrt{-a^2\mu^2}$, is shown in the
second column of Table \ref{table}, apart from the massless gauge
field associated to the residual $U(1)$.
\begin{figure}[t]
\begin{picture}(410,160)(0,120)

\Text(100,280)[]{{\bf Sphere ($SU(2)$ symmetry)}}
\BCirc(100,200){50} \Oval(100,200)(15,50)(0)

\Text(190,200)[]{ $\rightarrow$} \Text(300,280)[]{{\bf Ellipsoid
($U(1)$ symmetry)}} \Oval(300,200)(30,70)(0)
\Oval(300,200)(15,70)(0)

\end{picture}
\caption{Electroweak symmetry breaking in the geometrical approach.}
\label{distortion}
\end{figure}
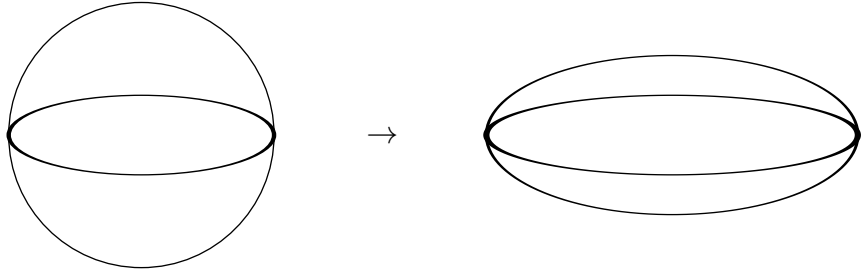
\paragraph{Geometrical Approach}

Now we want to compare the 4D effective theory with the geometrical
approach to SSB. By definition the latter involves a solution of the
higher dimensional EOMs that has the same symmetry as the effective
theory in the broken phase. At the leading non trivial order in the
small parameter $\eta^{1/2}\equiv \sqrt{-a^2\mu^2}$, we
find\footnote{This solution was discussed in Ref. \cite{S}, but
incorrectly.}
\begin{eqnarray} ds^2 &=&\eta_{\mu\nu}dx^{\mu}dx^{\nu}+ a^2\left[(1+|\eta|\beta
\sin^{2}\theta)d\theta^2+ \sin^2\theta d\varphi^2\right],\nonumber
\\
A&=&\frac{1}{e}(\cos\theta \mp 1)d\varphi,\label{solution2} \\
 \phi &= &\eta^{1/2}\alpha \exp\left(i\varphi \right) \sin\theta
 ,\nonumber
 \end{eqnarray}
where $\beta =\kappa^2 |\alpha|^2$ and $$|\alpha|^2=\frac{9}{32 \pi
a^4}\frac{1}{|\lambda_H-\lambda_G|}.$$ Consistency requires that if
$\mu^2>0$ then $\lambda_H<\lambda_G$, whereas if $\mu^2<0$ then
$\lambda_H>\lambda_G$. We are interested in $\mu^2<0$, as it
corresponds to the gauge symmetry breaking in the 4D effective
theory approach. Therefore we assume $\lambda_H>\lambda_G$. The VEV
of $\chi$, which corresponds to Solution (\ref{solution2}), is
$$
 |<\chi>|^2=\frac{3}{4}\,\,\frac{-\mu^2}{\lambda_H-\lambda_G}
$$
and we observe that it is equal to (\ref{VEV1}), apart from the
shift $\lambda_H\rightarrow \lambda_H-\lambda_G$. Moreover, the
metric that appears in Configuration (\ref{solution1}) describes an
$S^2$, whereas in (\ref{solution2}) we have the metric of an
ellipsoid. This distortion corresponds to the electroweak symmetry
breaking in the geometrical approach, as it is shown in Fig.
\ref{distortion}.
\begin{table}
\begin{tabular}{lrrrr}
\hline \tablehead{1}{l}{b}{Mass\\squared}
  &\tablehead{1}{r}{b}{}
  & \tablehead{1}{r}{b}{4D Effective \\Theory}
  &\tablehead{1}{r}{b}{}
  & \tablehead{1}{r}{b}{Geometrical \\Approach}
     \\
\hline $M^2_V$   & &$\frac{3e^2}{8\pi a^2}\frac{-\mu^2}{\lambda_H}$
& & $\frac{3e^2}{8\pi
a^2}\frac{-\mu^2}{\lambda_H-\lambda_G}$\\
& & & &\\
$M^2_{V_\pm}$   & &$\frac{9e^2}{16\pi a^2}\frac{-\mu^2}{\lambda_H}$ &&$\frac{9e^2}{16\pi a^2}\frac{-\mu^2}{\lambda_H-\lambda_G}$\\& & & &\\
$M^2_{F}$ & &$\frac{3g_Y^2}{16\pi a^2}\frac{-\mu^2}{\lambda_H}$
& & $\frac{3g_Y^2}{16\pi a^2}\frac{-\mu^2}{\lambda_H-\lambda_G}$ \\& & & &\\
$M^2_{F\pm}$ && $0$  &&  $0$ \\ & &&&\\$M^2_{S}$  & &$-2\mu^2$  & &$-2\mu^2$ \\& & & &\\
$M^2_{S\pm}$
&&$-\mu^2$&&$-\mu^2\frac{\lambda_H+\lambda_G}{\lambda_H-\lambda_G}$\\\hline
\end{tabular}
\caption{The vector (V), fermion (F) and scalar (S) spectra.}
\label{table}
\end{table}

The low energy vector, fermion and scalar spectrum\footnote{This
spectrum has been computed by using the formalism presented in Ref.
\cite{RS}.}, which corresponds to Solution (\ref{solution2}) is
presented in the third column of Table \ref{table}, apart from the
massless gauge field. We observe that, for vectors and fermions, the
only difference between the 4D effective theory and the geometrical
approach is the shift $\lambda_H\rightarrow \lambda_H-\lambda_G,$ as
for the VEV of $\chi$. However, concerning the scalar spectrum, we
have $M^2_S/M^2_{S\pm}=2,$ in the 4D effective theory approach,
whereas $M^2_S/M^2_{S\pm}=2(1-\delta)/(1+\delta)$, where
$\delta\equiv \lambda_G/\lambda_H,$ in the geometrical approach.
Since a ratio of masses is a measurable quantity, there is a
physical disagreement between the two approaches. The error is
measured by $\lambda_G/\lambda_H$ and we can roughly estimate its
magnitude: if we require $g_1$ and $g_2$ in (\ref{g12}) to be of the
order of $1$, and we also consider the relation between $\kappa$ and
the 4D Planck length, we obtain that $\lambda_G$ is of order of 1.
Therefore the condition $\lambda_G/\lambda_H\ll 1$ becomes
$\lambda_H\gg 1$, which is a strong coupling regime. Probably this
range is not allowed if one requires to study the 4D effective
theory by using perturbation theory. We conclude that the heavy mode
contribution to the low energy dynamics is in general non negligible
even in standard KK theories, where the heavy mode masses are
naturally at the Planck scale. Finally, we observe that this
contribution can be interpreted in a geometrical way, as the
internal space deformation of the 6D solution: indeed, if we put
$\beta=0$ but we keep $\alpha \neq 0$ in (\ref{solution2}), which
corresponds to neglecting the $S^2$ deformation, the spectra in
Table \ref{table} turn out to be equal.

\section{Non-Abelian and Supersymmetric Extensions}

\paragraph{Gauge-Higgs Unification} This scenario consists of models without fundamental scalars,
which, in some sense, geometrize the Higgs mechanism. Explicit
realizations, which include dynamical gravity, are presented in Ref.
\cite{Dvali:2001qr}. In particular, the authors analyzed a 6D
Einstein-Yang-Mills model, which is a non-Abelian extension, without
bulk scalars, of our theory. In a simple set up the bulk gauge group
is chosen to be $SU(3)$ and a non-Abelian generalization of Solution
(\ref{solution1}) can break $SU(3)$ down to $SU(2)\times U(1)$. The
internal components of the bulk gauge fields contain a doublet of
$SU(2)$, which can be naturally interpreted as a Higgs field. In
this way the Higgs mass is protected from dangerous power-law
radiative corrections by the bulk gauge symmetry.

 In the 4D effective theory approach the Higgs
doublet triggers the SSB of $SU(2)\times U(1)$ down to the
electromagnetic $U(1)$. The results presented in the present paper
and in \cite{Randjbar-Daemi:2006gf} suggest that this method
provides the correct 6D predictions for the observable quantities.
This is because, in our model, the solution of the EOMs of the 4D
effective theory can be lifted back to a solution of the complete 6D
theory, if the heavy modes are properly taken into account.

\paragraph{6D Supergravities} Other extensions of our work can be done in the context of supersymmetric versions of 6D gauge and
gravitational theories. In particular, 6D gauged\footnote{"gauged"
means that a subgroup of the R-symmetry group is promoted to a local
symmetry.} supergravities have attracted much interest for several
reasons. One of them is that the flat 6D space-time {\it is not} a
solution of the corresponding EOMs and the most symmetric solution
is $(Minkowski)_4\times S^2$, which has been shown recently to be
the {\it unique} maximally symmetric solution of such models
\cite{GGP}. Therefore, these theories provide a theoretical
explanation for the background that we have considered in the
previous section.

Moreover, 6D gauged supergravity compactifications share some
properties with 10D supergravity compactifications, whilst remaining
relatively simple, and so it can be used as a toy model for 10D
string theory compactifications \cite{toykklt}, in particular they
can give rise to chiral fermions in 4D. Furthermore, like in string
theory, the requirement of anomaly freedom is a strong guiding
principle to construct consistent models. Indeed the minimal version
of such gauged supergravity, {\it the Salam-Sezgin model}
\cite{Salam:1984cj}, suffers from the breakdown of local symmetries
due to the presence of gravitational, gauge and mixed anomalies,
which render this model inconsistent at the quantum level
\cite{Alvarez-Gaume:1983ig}; but it can be transformed in an anomaly
free model by choosing the gauge group and the supermultiplet in a
suitable way \cite{Randjbar-Daemi:1985wc}. Therefore, the extension
of our analysis to this context could be a first step towards the
study of the heavy modes in string theory compactifications.

Moreover, such 6D supergravities have been recently investigated in
connection with attempts to find a solution to the
 cosmological dark energy problem \cite{Burgess:2004ib}.
 Some 3-branes
 solutions and their perturbations, which can be relevant for this scenario, have been
 studied
 in Refs. \cite{GGP,GGP2}. These backgrounds are deformations of Background (\ref{solution1}),
 like our ellipsoid solution in the geometrical approach, but
 involving a warp factor and conical defects. This similarity suggests that our computation
 can be extended to 6D gauged supergravities expanded around these 3-brane solutions.
 If the heavy mode contribution
is physically relevant, the underlying 6D physics should manifest
itself in the low energy dynamics.

\begin{theacknowledgments}
  This work was supported in part by the Swiss
Science Foundation. We are appreciative of the hospitality at the
IPT of Lausanne and the support by INFN. We would also like to thank
S. Randjbar-Daemi, M. Shaposhnikov and K. Zuleta for illuminating
and stimulating discussions.
\end{theacknowledgments}

\bibliographystyle{aipproc}

\end{document}